\pdfoutput=1
\documentclass[
preprint,
titlepage,
nofootinbib,
amsmath,amssymb,
showkeys,
aps,
prd,
floatfix
]{revtex4-1}

\usepackage{graphicx}
\usepackage{bm}
\usepackage[utf8]{inputenc}
\usepackage[cal=boondoxo]{mathalfa}
\usepackage[ruled, linesnumbered]{algorithm2e}
\usepackage[english]{babel}
\usepackage{mathtools}
\usepackage{enumitem}
\usepackage{tabularx}

\usepackage{hyphenat}

\usepackage{booktabs}

\usepackage{amsthm}
\newtheorem{theorem}{Theorem}
\newtheorem{definition}{Definition}

\usepackage{hyperref}

\begin{document}

\title{Covariant constructive gravity: \\ A step-by-step guide towards alternative theories of gravity}

\author{Nils Alex}
\email{nils.alex@fau.de}
\author{Tobias Reinhart}
\email{tobi.reinhart@fau.de}
\affiliation{Department Physik, Friedrich-Alexander-Universit{\"a}t Erlangen-N{\"u}rnberg,\\
Staudtstr.~7, 91058 Erlangen, Germany}

\begin{abstract}
We propose a step-by-step manual for the construction of alternative theories of gravity, perturbatively as well as nonperturbatively. The construction is guided by no more than two fundamental principles that we impose on the gravitational dynamics: invariance under spacetime diffeomorphisms and causal compatibility with given matter dynamics, provided that spacetime is additionally endowed with matter fields. The developed framework then guides the computation of the most general, alternative theory of gravity that is consistent with the two fundamental requirements.
Utilizing this framework we recover the cosmological sector of general relativity solely from assuming that spacetime is a spatially homogeneous and isotropic metric manifold. On top of that, we explicitly test the perturbative framework by deriving the most general third-order expansion of a metric theory of gravity that is causally compatible with a Klein-Gordon scalar field. Thereby we recover the perturbative expansion of general relativity. To demonstrate how new physics emerges from our approach, we finally construct the most comprehensive third-order expansion of a theory of gravity that supports general (not necessarily Maxwellian) linear electrodynamics.
\end{abstract}


\maketitle

\section{Introduction}
Research on alternative theories of gravity faces a fundamental problem: While the experimental data consistently shows that the current description of the universe given by general relativity (GR) and the Standard Model of particle physics (SMPP) is incomplete, it is not obvious how a (more) complete theory should look like. Consider, for example, the problem of flat galaxy rotation curves \cite{Freeman_1970,Rubin_1970,Rubin_1980}. Most explanations for this phenomenon introduce dark matter \cite{Trimble_1987}, i.e., augment the SMPP with additional matter fields. There is, however, no consensus about the nature of these fields \cite{Bertone_2005}. In addition, this also leaves open the question whether GR has to be modified as well and, if so, how it must be modified. This leaves the researcher with an infinity of possible models to choose from. Testing such an infinity of models against experimental data cannot be done with only a finite number of astrophysical measurements. In spite of this impossibility, numerous proposed alternatives to GR \cite{Yunes_2013,Starobinsky_2007} tried to resolve the inconsistencies within the current models, but a real breakthrough has still not been achieved. To facilitate these undertakings, we propose that the problem of constructing alternative theories of gravity should best be tackled with a more structured plan at hand.

We call this plan \emph{covariant constructive gravity}. At the heart of it, there lie two fundamental requirements we impose on any alternative theory of gravity:
\begin{enumerate}[label=(A\arabic*)]
  \item The dynamical laws that govern gravity are invariant under spacetime diffeomorphisms. \label{axioms_intro_1}
  \item Provided that spacetime is additionally inhabited by matter fields, their dynamics is causally compatible with the gravitational dynamics. \label{axioms_intro_2}
\end{enumerate}
In the present work, we cast these two axioms in rigorous mathematical language and develop a procedure to derive the most comprehensive dynamical theory of gravity for a tensor field that is consistent with \ref{axioms_intro_1} and \ref{axioms_intro_2}. The framework can be generalized at will, for instance to also include nontensorial gravitational fields.

If in the model at hand spacetime is inhabited by matter fields with known dynamics, such that \ref{axioms_intro_2} has to be implemented, our approach is closely related to the program of (canonical) gravitational closure \cite{Giesel_2012,Schuller_2014,D_ll_2018}, which aims to close the joint model of matter and gravitational fields by deducing dynamical laws for the gravitational field as well. The difference, however, lies in the implementation of the axioms: While canonical gravitational closure works within the Hamiltonian picture, where general covariance is encoded in the constraint algebra, we choose the manifestly covariant Lagrangian description.

We proceed as follows: At the beginning of Sec.~\ref{chapter1}, we develop the necessary tools to obtain a rigorous, yet sufficiently general formulation of gravity as a second-derivative-order Lagrangian field theory in terms of the jet bundle formalism. This allows the two fundamental requirements to be cast in precise mathematical language: \ref{axioms_intro_1} is equivalent to a specific linear, first-order system of partial differential equations for the Lagrangian, whereas \ref{axioms_intro_2} can be formulated as additional algebraic conditions on the gravitational principal polynomial, a certain part of the gravitational equations of motion. Finishing Sec.~\ref{chapter1} we collect the results in the form of a precise manual for the construction of gravitational theories.

In Sec.~\ref{chapter2} we deduce perturbative equivalents to the two fundamental requirements that we formulated in Sec.~\ref{chapter1}. To that end, we concern ourselves with the computation of formal power series solutions to the PDE that we derived from \ref{axioms_intro_1}. This further allows us to extract viable information about how many independent curvature invariants any specific tensorial theory of gravity admits. In particular, we recover the known number of 14 curvature invariants for a spacetime endowed with a metric. We conclude Sec.~\ref{chapter2} with a concrete, algorithmic recipe for the perturbative construction of alternative theories of gravity. 

Section \ref{chapter4} is dedicated to testing the developed framework. In the nonperturbative setting we consider spacetime as a spatially homogeneous and isotropic metric manifold. Following along the steps of the construction manual, from this assumption we derive the Friedmann equations without ever using the Einstein-Hilbert Lagrangian from general relativity.
As a first successful test of the perturbative framework, we recover the perturbative expansion of general relativity from feeding the construction recipe the information that gravity be described by a metric tensor field.
Finally, we demonstrate how a modified theory of gravity may be obtained using our approach by constructing a third-order expansion of area metric gravity, the most general theory of gravity that is consistent with any linear theory of electrodynamics.

\section{The axioms of constructive gravity}\label{chapter1}
\subsection{Diffeomorphism invariant gravitational dynamics}

We wish to describe the gravitational field as a tensor field\footnote{The framework can readily be generalized to any vector bundle---with the restriction that for non-natural bundles, the lift of the diffeomorphism action from the base space to the total space has to be specified by hand.} over the 4-dimensional spacetime manifold $M$. To that end, let $F \subset T^m_nM$ be a vector subbundle of the $(m,n)$ tensor bundle over $M$, such that the gravitational field can be described as a section $G \in \Gamma(F)$ of this bundle. 
We denote adapted coordinates\footnote{We will always restrict to coordinates linear on the fibers \cite{Saunders_1989}.} on $F$ by $(x^m,v_A)$, where we introduced the abstract index $A$ that consequently runs over the fiber dimension of $F$.

As $F$ is a vector bundle, we define its vector bundle dual $F^{\ast}$ with fiber at $p\in M$ given by the vector space dual of $\pi_F^{-1}(p)$.
Moreover, we denote fiber coordinates dual to $v_A$ by $v^A$.

We might very well consider the case where $F$ represents a true subbundle of $T^m_nM$ and hence admits fibers of dimension $r < m+n$. For such situations it is convenient to introduce vector bundle morphisms that relate fiber coordinates $v_A$ on $F$ to fiber coordinates $v^{a_1 \dots a_m}_{b_1 \dots b_n}$ on $T^m_nM$.

\begin{definition}[intertwiner]\label{interDef}
Let $(F,\pi_F,M)$ be a vector bundle. We call a pair of vector bundle morphisms $(I, J)$,
\begin{equation}
    \begin{aligned}
    I&\colon F \rightarrow T^m_n M,\\
    J&\colon T^m_n M \rightarrow F,
    \end{aligned}
\end{equation}
that cover $id_M$ and satisfy
$J \circ I = \mathrm{id}_F$ a pair of intertwiners for the bundle $(F, \pi_F, M)$.
\end{definition}
In adapted coordinates we thus have the relations
\begin{equation} \label{interRel}
    \begin{aligned}
      v^{a_1 \dots a_m}_{b_1 \dots b_n} = {} & I^{A a_1 \dots a_m}_{b_1 \dots b_n} \cdot v_{A},\\  
      v_A = {} & J^{b_1 \dots b_n}_{A a_1 \dots a_m} \cdot v^{a_1 \dots a_m}_{b_1 \dots b_n},\\
      v^{b_1 \dots b_n}_{a_1 \dots a_m} = {} & J^{b_1 \dots b_n}_{A a_1 \dots a_m} \cdot v^{A},\\  
      v^A = {} & I^{A a_1 \dots a_m}_{b_1 \dots b_n} \cdot v^{b_1 \dots b_n}_{a_1 \dots a_m},\\
      \delta^A_B = {} & I^{A a_1 \dots a_m}_{b_1 \dots b_n} \cdot J^{b_1 \dots b_n}_{B a_1 \dots a_m}.  
    \end{aligned}
\end{equation}

The dynamics of the gravitational field shall be encoded as equations of motion to a second-derivative-order \emph{Lagrangian} for the gravitational field. We deliberately restrict to second-derivative-order Lagrangians, as any contribution of higher order necessarily leads to instabilities in the associated Hamiltonian formulation \cite{Ostrogradsky_1850}.

Such a gravitational Lagrangian can be rigorously defined on the \emph{jet bundle} \cite{Saunders_1989,Seiler_1994,Seiler_2010,Gotay_1992,Gotay_1998}.
We denote adapted coordinates of the second-order jet bundle $J^2F$ over $F$ by $(x^m, v_A, v_{Ap}, v_{AI})$. Here we introduced a new type of abstract index that is used to label second-order spacetime derivatives and thus runs from $0$ to $9$. 
The relation to the spacetime derivatives in standard notation is provided by an additional pair of intertwiners for the symmetric bundle $S_2M\subset T^0_2M$,
\begin{equation}
    \begin{aligned}
      v_{AI} = {} & J_I^{ij} v_{Aij},\\
      v_{Aij} = {} & I^I_{ij} v_{AI}.
    \end{aligned}
\end{equation}
We now define a second-derivative-order gravitational Lagrangian as follows.
\begin{definition}[Lagrangian]
A second-order Lagrangian $\mathcal L$ on $(F,\pi_F,M)$ is a bundle map that covers $id_M$:
\begin{equation}
    \mathcal{L} : J^2F \rightarrow \Lambda^4M.
\end{equation}
\end{definition}
Thus, the formulation of classical Lagrangian field theory yields the following situation, which is also illustrated in Fig.~\ref{diagram1}:
\begin{figure}[htp]
\centering
  \includegraphics{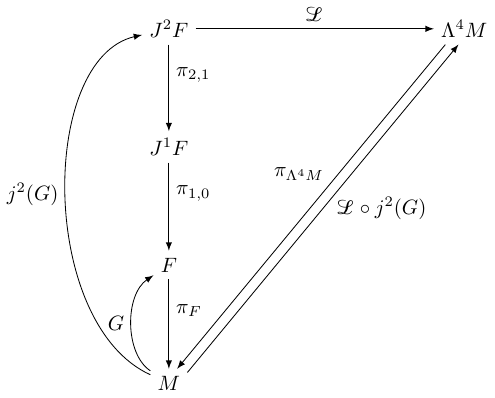}
\caption{Commutative diagram: Lagrangian field theory.} \label{diagram1}
\end{figure}
The gravitational field is described as a section of a bundle $F$ over the spacetime manifold $M$. As such it can be prolonged to any jet bundle $J^qF$, constructed over $F$, by applying the \emph{jet prolongation} map $j^q$. 
The Lagrangian $\mathcal{L}$ is a volume-form-valued bundle map on $J^2F$. Therefore, once composed with a prolonged section $j^2G$, we can compute its integral.
This defines the usual \emph{local action functional} on the space of fields:
\begin{equation}
\begin{aligned}
    S_{\mathcal{L}} \colon \Gamma(F) &\rightarrow \mathbb{R} \\
    G &\mapsto S_{\mathcal{L}}[G] \vcentcolon= \int \mathcal{L}(j^2(G)).
\end{aligned}
\end{equation}
Equations of motion (EOM) are obtained by equating the \emph{variational derivative} of the Lagrangian with zero:
\begin{equation}
\begin{aligned}
  0 = {} & E^A \\
    = {} & \frac{\delta \mathcal{L}}{\delta v_A} \\
    = {} & \frac{\partial\mathcal{L}}{\partial v_A} - D_p \left( \frac{\partial \mathcal{L}}{\partial v_{Ap}}\right) 
  + D_p D_q J^{pq}_I \left(\frac{\partial \mathcal{L}}{\partial v_{AI}}\right).
\end{aligned}
\end{equation}
Here we further introduced the jet bundle \emph{total derivative} $D_p$ that
maps a function $f$ on $J^qF$ to a function on $J^{q+1}F$:
\begin{equation}
    D_p f \vcentcolon= \frac{\partial f}{\partial x^p} + v_{Ap} \cdot  \frac{\partial f}{\partial v_A} + v_{AI} I^{I}_{pq} \cdot \frac{\partial f}{ \partial v_{Aq}}+\cdots .
\end{equation}
Note that the EOM of a second-order Lagrangian are thus, in general, given by a function on $J^4F$. As we wish to restrict to theories that allow for a meaningful Hamiltonian formulation, we will restrict, however, to those cases where $\mathcal{L}$ is \emph{degenerate}, s.t.~the EOM are also of second derivative order. 

\begin{figure}[t]
\centering
  \includegraphics{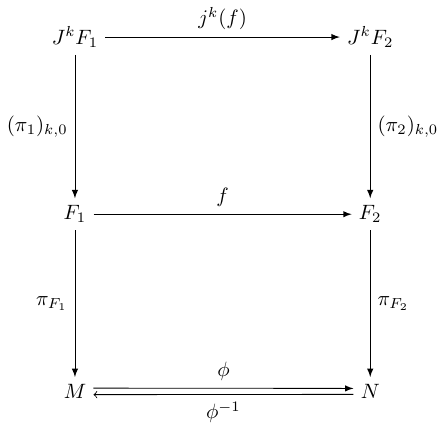}
\caption{Commutative diagram: prolongation of morphisms.} \label{diagram2}
\end{figure}

One of the fundamental requirements that we wish to impose on the gravitational dynamics is their \emph{invariance under spacetime diffeomorphisms}. 
This can be understood as a consequence of Einstein's requirement of general covariance \cite{Einstein_1916,Norton_1993}.
To that end, it is necessary that we lift the standard action of $\mathrm{Diff}(M)$ to $J^2F$.
As $F\subset T^m_nM$, the action of $\mathrm{Diff}(M)$ lifts naturally, by the usual pushforward-pullback construction, to an action by vector bundle isomorphisms on $F$.
In the following, we denote the image of $\phi \in \mathrm{Diff}(M)$ under this lift by $\phi_F$.
In order to further lift this action to the jet bundle we need to introduce an additional technique.
\begin{definition}[prolongation of morphisms]
Let $(F_1$,$\pi_{F_1}$,$M)$ and $(F_2,\pi_{F_2},N)$ be bundles, $\phi \colon M \rightarrow N$ a diffeomorphism, $f \colon F_1 \rightarrow F_2$ a bundle morphism covering $\phi$.
The $k$th-order jet bundle lift of $(f,\phi)$ is the unique map $j^k(f):J^kF_1 \rightarrow J^kF_2$ that lets the diagram in Fig.~\ref{diagram2} commute.
\end{definition}
Note that when acting on sections $G \in \Gamma(F_1)$, the jet bundle lift of bundle morphisms commutes with the jet prolongation map
\begin{equation}
j^k(f) \circ j^kG \circ \phi^{-1} = j^k \left (
f \circ G \circ \phi^{-1} \right ).
\end{equation}

Using this notion of lifted bundle morphisms we finally formulate the first fundamental requirement of constructive gravity in rigorous fashion. 
\begin{definition}
A Lagrangian field theory described by a second-order Lagrangian $\mathcal{L} \colon J^2F \rightarrow \Lambda^4 M$ is called diffeomorphism invariant if $\mathcal{L}$ is equivariant w.r.t.~the lifted action of $\mathrm{Diff}(M)$ on $J^2F$ and the pullback action on $\Lambda^4M$, i.e., if it holds for all $\phi \in \mathrm{Diff}(M)$ that
\begin{equation}\label{DiffeoReq}\tag{Axiom 1}
     \mathcal{L}\circ j^2(\phi_F) = \phi_{\ast} \circ \mathcal{L}.
\end{equation}
\end{definition}

Infinitesimally, on the \emph{Lie algebra level}, diffeomorphisms are described by vector fields $\Gamma(TM)$ with lie bracket provided by their commutator. The lifted action of $\mathrm{Diff}(M)$ on $F$ yields a Lie algebra morphism \cite{Gotay_1992}
\begin{equation}\label{LieF}
\begin{aligned}
  \mathcal{f} \colon \Gamma(TM) \rightarrow {} & \Gamma(TF) \\
    \xi \mapsto {} & \xi_F = \xi^m \frac{\partial}{\partial x^m} + \xi^A \frac{\partial}{\partial v_A} \\
    & \hphantom{\xi_F} = \xi^m \frac{\partial}{\partial x^m} + C_{An}^{Bm} v_B \partial_m \xi ^n \frac{\partial}{\partial v_A}. 
\end{aligned}
\end{equation}
Here we introduced the constants $C^{Bm}_{An}$ that constitute the vertical coefficient of this lifted vector field.
Further, we get the following Lie algebra morphism that yields the corresponding vector field on $J^2F$:
\begin{equation}
    \begin{aligned}
      j^2(\mathcal{f}) \colon \Gamma(TM) \rightarrow {} & \Gamma(TJ^2F)\\
      \xi \mapsto {} & \xi_{J^2F},
    \end{aligned}
\end{equation}
where 
\begin{equation}\label{LieJ2}
\begin{aligned}
  \xi_{J^2F} = {} &
     \xi^m \frac{\partial}{\partial x^m} + C_{An}^{Bm} v_B \partial_m \xi ^n \frac{\partial}{\partial v_A} \\
    & + C_{An}^{Bm} \partial_m \xi^n v_{Bi} \frac{\partial}{\partial v_{Ai}} - v_{An} \partial_m \xi ^n \frac{\partial}{\partial v_{Am}} \\
    & + C_{An}^{Bm} v_B \partial_m \partial_p \xi^n \frac{\partial}{\partial v_{Ap}} + C_{An}^{Bm} v_{BI} \partial_m \xi ^n \frac{\partial}{\partial v_{AI}} \\
    & - 2 v_{AJ} I^J_{an}J^{am}_I \partial_m \xi^n \frac{\partial}{\partial v_{AI}} +  2 C_{An}^{Bm} v_{Ba}J^{ap}_I \partial_m \partial_p \xi^n \frac{\partial}{\partial v_{AI}} \\
    & - v_{An} J^{pm}_I \partial_m \partial_p \xi^n\frac{\partial}{\partial v_{AI}} + C_{An}^{Bm} v_B J^{pq}_I \partial_m \partial_p \partial_q \xi^n \frac{\partial}{\partial v_{AI}}.
\end{aligned}
\end{equation}
We now use the Lie algebra morphism (\ref{LieJ2}) to derive an infinitesimal version of the first fundamental requirement (\ref{DiffeoReq}).
\begin{theorem}
Let $\mathcal{L} = L \mathrm{d}^4x$ be the Lagrangian of a diffeomorphism invariant field theory on $J^2F$, i.e., $\mathcal{L}$ is assumed to satisfy condition (\ref{DiffeoReq}). Then the coordinate expression $L$ necessarily satisfies the following system of first-order, linear partial differential equations,  
\begin{equation}\label{DiffeoEqn}
\begin{aligned}
  0 = {} & L_{:m} \\
    0 = {} & L^{:A} C_{An}^{Bm} v_B + L^{:Ap} \bigl[ C_{An}^{Bm} \delta_p^q - \delta_A^B \delta^q_n \delta^m_p  \bigr] v_{Bq}  + L^{:AI} \bigl[ C_{An}^{Bm} \delta_I^J - 2 \delta_A^B J_I^{pm} I^J_{pn}  \bigr] v_{BJ} + L \delta^m_n \\
    0 = {} & L^{:A(p\vert}C_{An}^{B \vert m)} v_B + L^{: AI} \bigl[ C_{An}^{B(m} 2 J_I^{p) q} - \delta^B_A J_I ^{pm} \delta_n^q \bigr] v_{Bq} \\
    0 = {} & L^{:AI} C_{An}^{B(m} v_B J_I^{p q )},
\end{aligned}
\end{equation}
where $L_{:m} \vcentcolon= \frac{\partial L}{\partial x^m}$, $L^{:A} \vcentcolon= \frac{\partial L}{ \partial v_A}$, etc.
\end{theorem}
\begin{proof}
Expressing condition (\ref{DiffeoReq}) infinitesimally, by utilizing the Lie algebra morphism (\ref{LieJ2}) for an arbitrary vector field $\xi \in \Gamma(TM)$, yields an equation with left-hand side given by applying $\xi_{J^2F}$ on $L$ and right-hand side given by the infinitesimal of the pullback action of $\phi$ on $\Lambda^4M$.
As $\xi \in \Gamma(TM)$ was assumed arbitrary, we can choose the individual components s.t.~specific contributions to the equations are isolated. These then have to be satisfied independently. Several suitable choices for the vector field components then yield precisely PDE system (\ref{DiffeoEqn}).
\end{proof}

This system of 140 first-order, linear partial differential equations for the Lagrangian follows necessarily from the requirement of diffeomorphism invariance. 
Conversely, every solution to this PDE yields a valid candidate Lagrangian to describe the gravitational dynamics. 
The problem of constructing gravitational dynamics is thus rephrased as computing the general solution to (\ref{DiffeoEqn}). 
This is an enormous advantage: As solving partial differential equations is a frequently occurring problem in almost all areas of research, the underlying theory is extensively developed \cite{Clebsch_1866,Courant_1937,Seiler_2010}.
Furthermore, note that the only quantity appearing in (\ref{DiffeoEqn}) that explicitly depends on the specific gravitational field is the vertical coefficient $C^{Bm}_{An}$. This allows for a unified treatment of the PDE, irrespective of the precise gravitational field at hand. 

Gotay \emph{et al.}~derived a similar system for first-order Lagrangians \cite{Gotay_1992,Gotay_1998} and used this system to define a universal, conserved energy-momentum tensor as Noether current associated to $\mathrm{Diff}(M)$. 
Unfortunately, restricting to first-order Lagrangians for a description of gravity does not always suffice---for instance, it does not in GR with only the usual metric tensor as gravitational field.

There are many exciting implications of (\ref{DiffeoEqn}) that, however, go beyond the scope of this paper. We have already presented a framework for the construction of perturbative gravitational dynamics that thrives on consequences of (\ref{DiffeoEqn}) on the corresponding EOM \cite{Reinhart_2018}.
Utilizing the jet bundle formulation of Hamiltonian dynamics \cite{Gotay_2004} one can further show---at least for first-derivative-order theories---that the Hamiltonian associated to any diffeomorphism invariant Lagrangian field theory is necessarily given by a linear combination of 4 primary and 4 secondary constraints and thus vanishes weakly \cite{Reinhart_2019}.

It is worth noting that diffeomorphism invariance also constitutes the main guiding principle for three well-known approaches that achieved to recover Einstein's general relativity as the unique, second-derivative-order, metric theory of gravity. 
First of all, Lovelock showed by directly imposing the condition of diffeomorphism invariance on the EOM that they are uniquely given by the Einstein tensor \cite{Lovelock_1969,Lovelock_1971,Lovelock_1972}.
Hojman \emph{et al.}~derived the canonical formulation of general relativity by requiring the Hamiltonian to be fully constrained and the corresponding constraint algebra to resemble the algebra of hypersurface deformations \cite{Hojman_1976}.
Ultimately this is also related to diffeomorphism invariant dynamics \cite{Reinhart_2019,Bojowald_2010}.
Last but not least, contributions mainly due to Deser revealed how Einstein dynamics can be obtained from energy-momentum conservation of the gravitational dynamics \cite{Deser_1970,Padmanabhan_2008}.
Conversely, Gotay \emph{et al.}~showed that their universal energy-momentum tensor is conserved and, moreover, reproduces the well-known expression in the case of general relativity \cite{Gotay_1992}.
Therefore, all three approaches illustrate further how diffeomorphism invariance---incorporated from three distinct points of view---can be seen as one of the fundamental traits that distinguish general relativity and further serves as an excellent guiding principle for the construction of gravitational dynamics.

\subsection{Causal compatibility between matter and gravity}

In the last section, we considered the formulation of a bare gravitational theory. If we additionally endow spacetime with a matter field $\phi \in  \Gamma(F_{\text{mat}})$ that is coupled to the gravitational field, i.e., whose dynamics is governed by a first-order Lagrangian\footnote{In this definition $\oplus_M$ is the Whitney sum of fiber bundles over the common bases space $M$ \cite{Karoubi_1978}.}
\begin{equation}\label{matterL}
    \mathcal{L}_{\text{mat}} \colon F_\text{grav} \oplus_M J^1F_\text{mat} \rightarrow \Lambda^4M,
\end{equation}
we additionally have to ensure that the description of matter and gravity are \emph{causally compatible}.

The causal structure of a given second-order EOM $E^A=0$ is closely related to the behavior of wave-like solutions in the infinite frequency limit \cite{D_ll_2018}. We consider the WKB ansatz for the coordinate expression of a section $G_A \in \Gamma(F)$
\begin{equation}\label{waveAns}
    G_A(x^m) = \mathrm{Re}\left \{ \mathrm e^{\frac{\mathrm iS(x^m)}{\lambda}} \cdot   \bigl [ a_A(x^m) + \mathcal{O}(\lambda) \bigr ]\right \}.
\end{equation}
Evaluating the EOM at this section and taking the limit $\lambda \rightarrow 0$ one obtains in leading order
\begin{equation}
    \underbrace{\left ( \frac{\partial E^A }{\partial v_{BI}} \right ) J_{I}^{ab} k_a k_b}_{T^{AB}(k_a)} a_B(x^m) = 0,
\end{equation}
where $k_a = - \partial_aS(x^m)$ is the wave covector of the ansatz. The $r\times r$ matrix $T^{AB}(k_a)$ is called the \emph{principal symbol} of the EOM. If the wave ansatz (\ref{waveAns}) with wave covector $k_a$ shall be a nontrivial solution with $a_A \neq 0$ to the EOM, then, in particular, $T^{AB}(k_a)$ must not be injective. 

Requiring such a square matrix to be noninjective is, of course, equivalent to imposing the vanishing of its determinant. There is, however, a caveat that obstructs this straightforward approach. If the theory at hand features gauge symmetries, its principal symbol is necessarily noninjective, irrespective of the specific covector $k_a$. 
The reason for this lies in the fact that for a gauge symmetry with $s$-dimensional orbits, there exist $s$ independent coefficient functions $\chi_{(i)A}(k_a)$, for $i = 1,\dots,s$, that are gauge-equivalent to the trivial solution $a_A(x^m)=0$ and thus are contained in the kernel of the principal symbol. 
Consequently, if we wish to obtain at least one physically nontrivial solution with wave covector $k_a$ that does not vanish modulo gauge transformation, we need to require that the kernel of $T^{AB}(k^m)$ is at least $s+1$ dimensional. This is equivalent to imposing the vanishing of all order-$s$ subdeterminants, i.e., a vanishing order-$s$ adjugate matrix
\begin{equation}
    Q_{(A_1\dots A_s) (B_1\dots B_s)}(k_a) \vcentcolon= \frac{\partial^s (\operatorname{det}(T^{AB}(k_a)))}{\partial T^{A_1 B_1}(k_a) \dots \partial T^{A_s B_s}(k_a)}.
\end{equation}  
It can be shown that $Q_{(A_1\dots A_s) (B_1\dots B_s)}(k_a)$ is subject to the general form \cite{Itin_2009,D_ll_2018}
\begin{equation}
     Q_{(A_1\dots A_s) (B_1\dots B_s)}(k_a) = \epsilon^{\sigma_1\dots\sigma_s} \epsilon^{\tau_1\dots\tau_s} \Big\lbrack\prod_{i=1}^{s} \chi_{(\sigma_i)A_i}(k_a)\Big\rbrack \Big\lbrack\prod_{j=1}^{s} \chi_{(\tau_j)B_j}(k_a)\Big\rbrack \mathcal{P}(k_a),
\end{equation}
where $\mathcal{P}(k_a)$ is a homogeneous, order $2r-4s$ polynomial in the covector components $k_A$. We call this function the \emph{principal polynomial} of the EOM.
Hence, in the infinite frequency limit, for (\ref{waveAns}) to describe a physically nontrivial solution to the EOM, it is necessary for the corresponding wave covector $k_a$ to be a root of the principal polynomial $\mathcal{P}(k_a)$. 
Thereby, the principal polynomial encodes the complete information of the propagation of wavelike solutions with infinite frequency. In particular, it contains the information about which spacetime domains such waves might \emph{causally influence} \cite{Khavkine_2012,Seiler_2010,R_tzel_2011}.

For the special case of $E^A$ describing diffeomorphism invariant dynamics, it follows from (\ref{DiffeoEqn}) that the following 4 independent coefficient functions lie in the kernel  of $T^{AB}(k_a)$: 
\begin{equation}
   \chi_{(n)A}(k_a) =  C_{An}^{Cm}v_Ck_m.
\end{equation}
In addition to defining admissible wave covectors of nontrivial solutions, the principal polynomial also provides information about suitable \emph{initial data hypersurfaces} that can serve as starting point for the initial value formulation of the theory, provided such a formulation exists.
\begin{theorem}
If the Cauchy problem of a given PDE is well-posed in a region of $M$, then the principal polynomial necessarily restricts to a hyperbolic polynomial on $T_p^{\ast}M$ for every $p$ contained in that region. Furthermore, exactly those hypersurfaces that have at every point a conormal which is hyperbolic w.r.t.~$\mathcal{P}$ are admissible initial data hypersurfaces.
\end{theorem}
\begin{proof}
The proof can be found in \cite{H_rmander_1977} and also in \cite{Ivrii_1974}.
\end{proof}
Note that the existence of a well-defined Cauchy problem is of fundamental importance for any meaningful physical theory, as only then the theory admits \emph{predictive} power \cite{Hern_ndez_2012}.  Thus, in the following, we will restrict all considerations to theories that feature hyperbolic EOM.

\begin{figure}[htp]
\begin{minipage}{0.45\textwidth}
\begin{center}
  \includegraphics{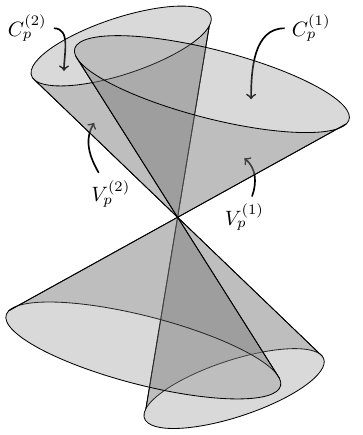}
\end{center}
\end{minipage}
\begin{minipage}{0.45\textwidth}
\begin{center}
  \includegraphics{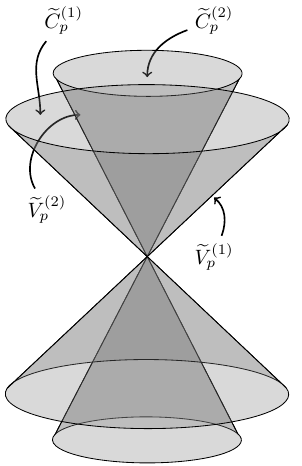}
\end{center}
\end{minipage}
    \caption{Vanishing sets $V_p$, $\widetilde{V}_p$ and hyperbolicity cones $C_p$, $\widetilde{C}_p$ of two hyperbolic polynomials of degree 4. In both cases the vanishing sets are given as union of the individual vanishing sets, i.e., the cone surfaces $V_p = V_p^{(1)} \cup V_p^{(2)}$, $\widetilde{V} = \widetilde{V}_p^{(1)} \cup \widetilde{V}_p^{(2)}$, while the hyperbolicity cones are provided by the intersection of the individual hyperbolicity cones $C_p = C_p^{(1)} \cap C_p^{(2)}$,  $\widetilde{C}_p = \widetilde{C}_p^{(1)} \cap \widetilde{C}_p^{(2)}$.}
    \label{Poly}
\end{figure}

Summing up, we see that the principal polynomial of any hyperbolic EOM defines in each $T_p^{\ast}M$, by means of its vanishing set $V_p\subset T^{\ast}_pM$, the set of admissible infinite frequency wave covectors. Moreover, the set of hyperbolic covectors $C_p \in T_p^{\ast}M$ that can be shown to constitute a convex cone \cite{G_rding_1959}, the so-called hyperbolicity cone, provides the relevant information of possible choices of initial data hypersurfaces. 
The situation is illustrated in Fig.~\ref{Poly}.

When constructing alternative theories of gravity, the distribution of vanishing sets $V_{\text{grav}} \subset T^{\ast}M$ of the gravitational principal polynomial $\mathcal{P}_{\text{grav}}$ thus not only governs the propagation of \emph{gravitational waves}, but its distribution of hyperbolicity cones $C_{\text{grav}}\subset T^{\ast}M$ further encodes suitable initial data hypersurfaces, which are closely related to viable \emph{observer definitions} \cite{D_ll_2018,R_tzel_2011,Hern_ndez_2012}. 

If gravity is additionally coupled to a matter field, we get a principal polynomial of the matter EOM $\mathcal{P}_{\text{mat}}$, which endows spacetime with an additional distribution of its vanishing sets $V_{\text{mat}}$ and its hyperbolicity cones $C_{\text{mat}}$. It is then crucial that the causal structures of matter and gravity are compatible. 
Not only do we have to impose $C_{\text{grav}} = C_{\text{mat}}$ if the two theories shall allow for a unified observer definition \cite{Hern_ndez_2012,R_tzel_2011}, recent observations of gravitational waves also suggest that they\footnote{At least those modes that have already been detected.} propagate at the speed of light \cite{Abbott_2017} and thus admit the same wave covectors as matter waves.
Therefore, we further need to require that wave covectors of the given matter theory also serve as wave covectors of gravitational waves: $V_{\text{mat}} \subset V_{\text{grav}}$.
Consequently, specifying any matter theory of the form (\ref{matterL}) that is coupled to gravity, we get two additional conditions on the gravitational dynamics:
\begin{equation}\label{A2}
    C_{\text{grav}} = C_{\text{mat}} \quad \text{and} \quad V_{\text{mat}} \subset V_{\text{grav}}.\tag{Axiom 2}
\end{equation}

Given such a matter theory we therefore proceed as displayed in Algorithm \ref{Algo1} in the construction of a compatible theory of gravity.
\begin{algorithm}[hbt]
\SetAlgoLined
\KwData{Matter theory $ \mathcal{L}_{\text{mat}} : F_{\text{grav}} \oplus_M J^1F_{\text{mat}} \rightarrow \Lambda^4M$.
}
\KwResult{Most general diffeomorphism invariant, causally compatible theory of gravity: $\mathcal{L}_{\text{grav}} : J^2F_{\text{grav}} \rightarrow \Lambda^4M$ .}
Compute $C^{Bm}_{An}$. \\
Set up PDE (\ref{DiffeoEqn}). \\
Solve PDE (\ref{DiffeoEqn}) for $\mathcal{L}_{\text{grav}}(x^m,v_A,v_{Am},v_{AI})$.\\
Compute $\frac{\delta \mathcal{L}_{\text{grav}}}{\delta v_A}$.\\
Restrict to $2$nd-derivative-order sub theory of $\mathcal{L}_{\text{grav}}$.\\
Calculate $\mathcal{P}_{\text{grav}}$ and $\mathcal{P}_{\text{mat}}$.\\
Impose $C_{\text{grav}} = C_{\text{mat}}$ and $V_{\text{grav}} \subset V_\text{mat}.$
 \caption{Construction of gravitational Lagrangian}\label{Algo1}
\end{algorithm}
\section{Perturbation theory}\label{chapter2}
\subsection{Perturbative diffeomorphism invariance}
Although (\ref{DiffeoEqn}) merely constitutes a linear, first-order PDE system---solving such systems is a well-studied subject \cite{Courant_1937}---practically computing the general solution poses a real problem for most cases. The reason is the sheer size of the system. Already when treating the relatively simple case of metric theories of gravity, Eq.~(\ref{DiffeoEqn}) consists of 140 partial differential equations for a function that depends on 154 independent variables.

Utilizing techniques from formal PDE theory to obtain solutions under special circumstances, the PDE system (\ref{DiffeoEqn}) nevertheless furnishes us with access to two significant realms of gravitational physics. First of all, we can employ methods of symmetry reduction and thereby obtain, for instance, the cosmological equivalent to the alternative theory of gravity in consideration. Such an approach is illustrated in Sec.~\ref{cosmo} and will be addressed in more detail in future work \cite{Alex_2020_PhD}.
Second---and this is the path that we pursue in the following---we can construct \emph{power series solutions} to (\ref{DiffeoEqn}) in order to retrieve a \emph{perturbative} description of the modified theory of gravity. Such a perturbative description of alternative theories of gravity in particular allows for the treatment of propagation and emission of \emph{gravitational waves}.
With the recent developments in the detection of gravitational waves, they provide an excellent test for modified theories of gravity \cite{Yagi_2010,Berry_2011,Zhang_2017,Yunes_2013}.

As first step in the construction of a power series solution to PDE (\ref{DiffeoEqn}), the Lagrangian is expanded around an expansion point $p_0 \in J^2F$. We choose an expansion point that serves as coordinate representation of a \emph{flat} instance of the gravitational field:
\begin{equation}
    J^2F \ni p_0 \equiv (x_0^m, N_A, 0, 0).
\end{equation}
Moreover, we want to incorporate the fact that, on small scales, spacetime is in reasonable approximation described by the geometry of Minkowski spacetime. We therefore choose $N_A=N_A(\eta_{ab})$ such that we can interpret the perturbative theory of gravity as an expansion around Minkowski spacetime, where $N_A$ are obtained as functions of the Minkowski metric $\eta_{ab}$, such that:
\begin{enumerate}[label=(\roman*)]
    \item $\mathcal{L}_{\text{mat}} (N_A, -)$ yields a description of the matter field equivalent to placing it on Minkowski spacetime. 
    \item $N_A(\eta_{ab})$ is \emph{Lorentz invariant}, i.e., it holds that $0 = N_A C^{Am}_{Bn}(K_{(i)})^n_m$ for the 6 Lorentz generators $(K_{(i)})^n_m \in \bigl \{\eta^{\vphantom{n}}_{m [r}\delta^n_{s]} \mid r < s \bigr \}$.
\end{enumerate}
We define the coordinate deviation from the expansion point,
\begin{equation}
    (H_A,H_{Ap},H_{AI}) = (v_A-N_A, v_{Ap}, v_{AI}),
\end{equation}
and expand the gravitational Lagrangian $L_{\text{grav}}$ as formal power series around $p_0$ up to finite order $q$:
\begin{equation}\label{LperRed}
\begin{aligned}
  L_{\text{grav}} = {} & a_0 + a^A H_A + a^{AI}H_{AI} \\
     & + a^{AB} H_{A}H_{B} + a^{ApBq} H_{Ap}H_{Bq} + a^{ABI} H_{A} H_{BI}\\
     & + a^{ABC} H_A H_B H_C + a^{ABpCq} H_{A}H_{Bp}H_{Cq} \\
     & + a^{ABCI} H_A H_B H_{CI} + \dots + \mathcal{O}(q+1).
\end{aligned}
\end{equation}
Here the expressions $a_0, a^{A},\dots$ are constants.
Note that we do not include any explicit dependency on $x^m$ in the expansion, as the first equation in (\ref{DiffeoEqn}) prohibits such. 
Further note that terms that include a total of more than two spacetime derivatives\footnote{For instance, $a^{ApBI}H_{Ap}H_{BI}$ includes a total of 3 spacetime derivatives.} must be removed from the expansion (\ref{LperRed}) as these terms would necessarily make the gravitational principal polynomial $\mathcal{P}_{\text{grav}}$ depend on derivative coordinates $v_{Ap}$ and $v_{AI}$ and thus be causally incompatible with $\mathcal{P}_{\text{mat}}$.
Finally, note that we excluded terms that only feature a single spacetime derivative, as such are prohibited for the following reason: We will subsequently show how the required Lorentz invariance of the expansion point $N_A$ enters the PDE (\ref{DiffeoEqn}) and imposes restrictions on the expansion coefficients. In particular, we will see that this forbids terms with a single spacetime derivative.

Inserting the expansion (\ref{LperRed}) into PDE (\ref{DiffeoEqn}) and evaluating at the expansion point $p_0$ yields the following system of linear equations for the first-order expansion coefficients:
\begin{equation}\label{order1}
    \begin{aligned}
      0 = {} & a^A C_{An}^{Bm}N_B + a_0 \delta^m_n \\
      0 = {} & a^{AI}C_{An}^{B(m}N_B J^{pq)}_I.
    \end{aligned}
\end{equation}
Prolonging the PDE to second derivative order, inserting the expansion and again evaluating at the expansion point, we obtain linear equations for the second-order expansion coefficients\footnote{Here, and below, $C_{An}^{m} = C_{An}^{Dm} N_D$.}: 
\begin{equation}\label{order2}
    \begin{aligned}
      0 = {} & a^A C_{An}^{Bm} + 2 a^{AB}C_{An}^{m} + a^B\delta^m_n\\
      0 = {} & a^{AI}\left [C_{An}^{Bm}\delta^J _I- 2 \delta^B_A J_I^{pm}I^J_{pn} \right ] + a^{ABJ}C_{An}^{m} + a^{BJ} \delta^m_n \\
      0 = {} & 2a^{A(p\vert Bq}C_{An}^{\vert m)} + a^{AI} \left [C_{An}^{B(m} 2 J_{I}^{p)q} - \delta_A^BJ_I^{pm}\delta^q_n \right ]\\
      0 = {} & a^{BAI}C_{An}^{(m}J_I^{pq)} + a^{AI}C_{An}^{B(m} J_I^{pq)}.
    \end{aligned}
\end{equation}
Proceeding analogously we derive further linear equations for the third-order expansion coefficients:
\begin{equation}\label{order3}
\begin{aligned}
  0 = {} & 2 a^{AC}C_{An}^{Bm} + 2a^{AB}C_{An}^{Cm} + 6 a^{ABC}C_{An}^{m} + 2a^{BC} \delta^m_n\\
  0 = {} & 2 a^{ApCr} \left [ C_{An}^{Bm} \delta ^q_p - \delta^B_A \delta^m_p\delta^q_n \right ] +2 a^{A Bq Cr} C_{An}^{m} + 2 a^{BqCr} \delta^m_n\\
  0 = {} & a^{CAI} \left [C_{An}^{Bm}\delta^J _I- 2 \delta^B_A J_I^{pm}I^J_{pn} \right ] + 2 a^{ACBJ} C_{An}^{m} + a^{CBJ} \delta ^m _n \\
  0 = {} & 2 a^{C A(p \vert B q} C_{An}^{\vert m )} + a^{CAI} \left [C_{An}^{B(m} 2 J_{I}^{p)q} - \delta_A^BJ_I^{pm}\delta^q_n \right ]\\
  0 = {} & 2 a^{BCAI}C_{An}^{(m}J_I^{pq)} + a^{CAI}C_{An}^{B(m} J_I^{pq)}.
\end{aligned}
\end{equation}
Similarly, one readily obtains the corresponding linear equations for any higher-order expansion coefficients. 

When constructing such power series solutions to a given PDE system, it is crucial to know whether or not the system generates \emph{integrability conditions} during prolongations. Such lower-derivative-order equations that are only present once the system is prolonged to a higher derivative order thoroughly disturb the perturbative treatment of the system. This is because perturbation theory is usually motivated by the fact that by aborting the construction of a power series solution
after some order $q>0$, the difference to the exact solution is of $\mathcal{O}(q+1)$ in the deviation from the expansion point. Thus, such a perturbative solution provides a reasonable approximation near $p_0$. 
If now during some higher order prolongations the PDE produces integrability conditions of derivative order lower than $q$, one gets additional equations that further restrict the computed perturbative solution. 
Hence, the perturbative solution actually does not approximate the real problem with the desired accuracy. Putting it differently, in such a case, one, in fact, misses information that is hidden in the integrability conditions and therefore obtains a solution that is too general, i.e., includes fake solutions. 

PDEs that are certain not to produce any integrability conditions are called formally integrable.
Proving formal integrability of PDEs is notoriously difficult. There exists, however, the related notion of \emph{involutive} PDEs that implies formal integrability, i.e., constitutes a stronger condition, whilst at the same time is much easier to verify \cite{Seiler_1994,Seiler_2010}.  
\begin{theorem}\label{Invol}
PDE system (\ref{DiffeoEqn}) is involutive and thus, in particular, formally integrable.
\end{theorem}
\begin{proof}
We only sketch why any potential integrability condition of (\ref{DiffeoEqn}) is necessarily linearly dependent on the equations already contained in (\ref{DiffeoEqn}) and thus provides no further information needed to construct a power series solution.
Details can be found in \cite{Seiler_1994} and \cite{Reinhart_2019}.
The claim follows from the fact that the homogeneous part of PDE system (\ref{DiffeoEqn}) is described by vector fields that span the image of the lie algebra morphism (\ref{LieJ2}) in $\Gamma(J^2F)$. The only way for linear PDEs to generate integrability conditions is by adding several prolongations of individual equations in the system such that all second-derivative-order contributions cancel due to the commutative law of second partial derivatives. The remaining first-derivative-order contributions can then be shown to be given by a commutator of vector fields contained in the image of (\ref{LieJ2}). As this image, in particular, constitutes a Lie algebra, this commutator is necessarily given as linear combination of individual vector fields and thus vanishes modulo PDE system (\ref{DiffeoEqn}). Hence, the first-order contribution is already contained in (\ref{DiffeoEqn}) and therefore does not constitute an integrability condition. 
\end{proof}
Theorem \ref{Invol} implies that the previously described perturbation techniques can safely be applied to PDE (\ref{DiffeoEqn}), without risking to obtain perturbative solutions that are too general. 

Involutive PDEs admit many unique properties. For instance, they allow for a straightforward prediction of the form of the general solution.
\begin{theorem}\label{FormalSol}
The general solution to (\ref{DiffeoEqn}) is of the form
\begin{equation}
    \omega \cdot \mathcal{F} \left (\Psi_1,\dots,\Psi_k \right ),
\end{equation}
where $k\vcentcolon= \operatorname{dim}(J^2F) - 140$,  $\Psi_1,\dots,\Psi_k$ are $k$ functionally independent solutions to the homogeneous PDE system corresponding to (\ref{DiffeoEqn}), $\mathcal{F}$ is any function of $k$ real variables, and $\omega$ is a particular solution to (\ref{DiffeoEqn}).
\end{theorem}
\begin{proof}
As proven in proposition 7.1 in \cite{Seiler_1994}, the general solution to the homogeneous PDE system corresponding to (\ref{DiffeoEqn}) is given by $\mathcal{F} \left (\Psi_1,\dots,\Psi_k \right )$. The claim can then readily be proven by noting that the product of any two solutions $\omega$ to (\ref{DiffeoEqn}) and $\rho$ to the homogeneous version is again a solution to PDE system (\ref{DiffeoEqn}) and conversely the quotient of two solutions $\omega_1, \omega_2$ to (\ref{DiffeoEqn}) solves the corresponding homogeneous version, which simply follows from the product rule of derivatives. \end{proof}

The functions $\Psi_i: J^2F \rightarrow \mathbb{R}$ are diffeomorphism invariant scalar functions. In the context of general relativity these are called \emph{curvature invariants}. It is a well-known result that the metric structure present in GR admits 14 functionally independent curvature invariants \cite{Narlikar_1949,Zakhary_1997}. 
Note that by the above theorem we readily recover this result. The metric fiber bundle $F_{\text{GR}}\vcentcolon=S_2M$ has fiber dimension 10, which admits 40 first-order derivative coordinates and 100 second-order derivative coordinates. Thus, we get $\operatorname{dim}(J^2F_{\text{GR}}) = 4+10+4\cdot10+10\cdot10 = 154$ and therefore $k=154-140=14$ curvature invariants. 
We are now, however, in a position to predict the number of independent curvature invariants for any other spacetime geometry at wish, simply by counting the dimension of the second-order jet bundle $\operatorname{dim}(J^2F)$.

When computing perturbative solutions to (\ref{DiffeoEqn}), there exists one further obstruction that is caused by the required Lorentz invariance of the flat expansion point $N_A$. Due to this additional symmetry, the second equation in (\ref{DiffeoEqn}) admits rank defects once evaluated at $p_0$. Consider this equation evaluated at a general point that admits a coordinate expression with vanishing derivative contributions $p \equiv (x^m,M_A,0,0)$:
\begin{equation}
0 = L^A \big \vert_{p} C_{An}^{Bm}M_B + a_0 \delta^m_n.
\end{equation}
This tensor equation in general contains 16 independent scalar equations. Conversely, evaluating the same equation at $p_0\equiv (x_0^m, N_A,0,0)$ yields only 10 independent equations, as contraction with the Lorentz generators $(K_{(i)})^n_m$ yields 6 independent vanishing linear combinations.\footnote{This contraction corresponds to the restriction from the diffeomorphism-induced local action of $\mathrm{GL}(4)$ in (\ref{DiffeoReq}) to the local action of the Lorentz group $\mathrm{SO}(3,1)$.}

When constructing a power series solution around a Lorentz invariant expansion point $p_0$, we can now form exactly the same linear combination for any prolongation of the second equation in (\ref{DiffeoEqn}). As the highest-derivative-order contribution of all these prolongations is proportional to $C^{Bm}_{An}N_B$, contracting with $(K_{(i)})^n_m$ always yields an additional equation of submaximal derivative order. The equations that we obtain by this procedure must, however, not be confused with integrability conditions, as they are only present once we evaluate at $p_0$. 

To provide an example for such an additional equation, we consider the first equation of (\ref{order2}) and contract against the Lorentz generators to obtain
\begin{equation}\label{ansatz1}
    0 = a^A C^{Bm}_{An}  (K_{(i)})^n_m.
\end{equation}
This additional equation for the first-order expansion coefficient $a^A$ has to be taken into account when constructing power series solutions to (\ref{DiffeoEqn}).
It states that also the expansion coefficient $a^A$ must be Lorentz invariant. 
Similar equations can be obtained from any further prolongation of the second equation in (\ref{DiffeoEqn}). These additional equations then also subject all remaining expansion coefficients to the invariance under Lorentz transformations.
We conclude that when computing power series solutions to (\ref{DiffeoEqn}) around a Lorentz invariant expansion point, the PDE itself demands that the expansion coefficients are also Lorentz invariant.

\subsection{Perturbative causal compatibility}
Once (\ref{DiffeoEqn}) is solved perturbatively, deducing the perturbative equivalent to the second axiom (\ref{A2}) is straightforward. We start by computing a perturbative expression of the matter principal polynomial, i.e., we simply expand $\mathcal{P}_{\text{mat}}$ around $p_0$ to obtain:
\begin{equation}
    \mathcal{P}_{\text{mat}} = (P_{\text{mat}}^{(0)}) + (P_{\text{mat}}^{(1)})^A H_A +\mathcal{O}(2).
\end{equation}
Here, we will stick to $L_\text{grav}$ expanded to third order, so it suffices to expand the two polynomials up to first order. All calculations are easy to generalize to higher orders.

The gravitational principal polynomial is obtained by first computing the perturbative EOM of the solution to (\ref{order1})--(\ref{order3}). This induces an expansion of the gravitational principal symbol that we denote as\footnote{We suppress matrix indices and any explicit $k_a$ dependency for the sake of a more concise notation.}
\begin{equation}
    T = (T^{(0)}) + (T^{(1)})^AH_A + \mathcal{O}(2).
\end{equation}
Moreover, we also expand
\begin{equation}
\begin{aligned}
  \chi_{(n)A} =\hphantom{\vcentcolon} {} &  C^{Bm}_{An} N_B k_m + C^{Bm}_{An} H_B k_m \\
  =\vcentcolon {} & (\chi^{(0)})_{An} + (\chi^{(1)})^B_{An}H_B
\end{aligned}
\end{equation}
and define
\begin{equation}\label{PreF}
f_{(A_1\dots A_4)(B_1\dots B_4)} \vcentcolon= \epsilon^{i_1\dots i_4} \epsilon^{j_1\dots j_4} \Big\lbrack\prod_{r=1}^{4} \chi_{(i_r)A_r}\Big\rbrack \Big\lbrack\prod_{s=1}^{4} \chi_{(j_s)B_s}\Big\rbrack,
\end{equation}
with the induced expansion
\begin{equation}
f_{(A_1\dots A_4)(B_1\dots B_4)} = (f^{(0)})_{(A_1\dots A_4)(B_1\dots B_4)} + (f^{(1)})^C_{(A_1\dots A_4)(B_1\dots B_4)}\,H_C + \mathcal{O}(2).
\end{equation}
We now choose a $(r-4) \times (r-4)$ full-rank submatrix $Q_{(A_1\dots A_4)(B_1\dots B_4)}$ of the principal symbol by removing appropriate rows $(A_1\dots A_4)$ and columns $(B_1\dots B_4)$. Its determinant expands as
\begin{equation}
    \operatorname{det}\bigl(Q_{(A_1\dots A_4)(B_1\dots B_4)})\bigr) = (D^{(0)})_{(A_1\dots A_4)(B_1\dots B_4)} + (D^{(1)})^C_{(A_1\dots A_4)(B_1\dots B_4)}\,H_C + \mathcal{O}(2),
\end{equation}
with the following contributions in the individual orders:
\begin{equation}\label{polyMatrices}
  \begin{aligned}
    (D^{(0)})_{(A_1\ldots A_4)(B_1\ldots B_4)}   = {} &  \operatorname{det}\bigl((Q^{(0)})_{(A_1\ldots A_4)(B_1\ldots B_4)}\bigr) \\
    (D^{(1)})^C_{(A_1\ldots A_4)(B_1\ldots B_4)} = {} & \operatorname{det}\bigl((Q^{(0)})_{(A_1\ldots A_4)(B_1\ldots B_4)}\bigr) \\
    & \times \operatorname{Tr} \bigl ( (Q^{(0)})^{-1}_{(A_1\ldots A_4)(B_1\ldots B_4)} 
   (Q^{(1)})_{(A_1\ldots A_4)(B_1\ldots B_4)}^C \bigr)
  \end{aligned}
\end{equation} 
This then finally allows us to expand the gravitational principal polynomial
\begin{equation}\label{POLYCoeffs}
    \mathcal{P}_{\text{grav}} = (P_{\text{grav}}^{(0)}) + (P_{\text{grav}}^{(1)})^A H_A + \mathcal{O}(2),
\end{equation}
where the individual contributions are given by
\begin{equation}
(P_{\text{grav}}^{(0)}) = \frac{(D^{(0)})_{(A_1\dots A_4)(B_1\dots B_4)}}{(f^{(0)})_{(A_1\dots A_4)(B_1\dots B_4)}}
\end{equation}
and
\begin{equation}
(P^{(1)}_{\text{grav}})^C = \frac{(D^{(1)})^C_{(A_1\ldots A_4)(B_1\ldots B_4)} - (f^{(1)})^C_{(A_1\ldots A_4)(B_1\ldots B_4)} \cdot (P_{\text{grav}}^{(0)})}{(f^{(0)})_{(A_1\ldots A_4)(B_1\ldots B_4)}}.
\end{equation}

Given the perturbative expressions for the matter and gravitational principal polynomial, we now implement the second axiom perturbatively by imposing that (\ref{A2}) holds in the corresponding order:
\begin{equation}\label{pertA2}
    C_{\text{grav}} = C_{\text{mat}} + \mathcal{O}(q-2) \quad \text{and} \quad V_{\text{mat}} \subset V_{\text{grav}} + \mathcal{O}(q-2).
\end{equation}

Summing up, given the perturbative expansion of the gravitational Lagrangian---which is necessarily of the form presented in (\ref{LperRed})---to the desired order around the chosen expansion point, the perturbative construction of alternative theories of gravity boils down to solving linear systems (\ref{order1})--(\ref{order3}) for the expansion coefficients. Doing so, one has to take into account that the expansion coefficients are necessarily Lorentz invariant if the expansion is chosen as such. Finally, condition (\ref{pertA2}) has to be imposed. The enormous advantage of the thus developed framework lies in the fact that the involved computations are not only conceptually entirely clear, but can easily be performed employing efficient computer algebra \cite{Reinhart_2019_sparse-tensor}, as they almost only involve basic linear algebra. Obtaining valid perturbative models of alternative theories of gravity is therefore merely a problem of setting up and solving the relevant equations. A precise step-by-step recipe for the computation of perturbative theories of gravity is displayed in Algorithm \ref{Algo2}.
\begin{algorithm}[hbt]
\SetAlgoLined
\KwData{Matter theory $\mathcal{L}_{\text{mat}} \colon F_{\text{grav}} \oplus_M J^1F_{\text{mat}} \rightarrow \Lambda^4M$, expansion order $q > 0$, Lorentz invariant expansion point $J^2F_{\text{grav}} \ni p_0 \equiv (x_0^m, N_A,0,0)$.}
\KwResult{Most general diffeomorphism invariant, causally compatible $\mathcal{L}_{\text{grav}}$ expanded as finite power series to order $q$ around $p_0$.}
Compute $C^{Bm}_{An}$. \\
Compute the most general Lorentz invariant expansion coefficients (use \cite{Reinhart_2019_sparse-tensor}).\\
Set up equations (\ref{order1})--(\ref{order3}) and all necessary higher-order equivalents.\\
Solve this linear system (use \cite{Reinhart_2019_sparse-tensor}).\\
Compute the expansion of $T_{\text{grav}}$.\\
Choose a full-ranked $Q_{(A_1\dots A_4)(B_1\dots B_4)}$. \\
Compute (\ref{PreF}).\\
Compute the expansion of $\operatorname{det}(Q_{(A_1\dots A_4)(B_1\dots B_4)})$ (\ref{polyMatrices}). \\
Compute the expansion $\mathcal{P}_{\text{grav}}$ (\ref{POLYCoeffs}). \\
Compute $\mathcal{P}_{\text{mat}}$ up to $\mathcal{O}(q-2)$.\\
Impose $C_{\text{grav}} = C_{\text{mat}} + \mathcal{O}(q-2) \ \ \text{and} \ \ V_{\text{mat}} \subset V_{\text{grav}} + \mathcal{O}(q-2)$.
 \caption{Perturbative construction of gravitational Lagrangian}\label{Algo2}
\end{algorithm}

\section{Applications}\label{chapter4}
After presenting the framework in general, we are now going to employ the construction recipes \ref{Algo1} and \ref{Algo2}
to derive three particularly interesting cases of gravitational dynamics, one constituting an exact theory and two perturbatively expanded theories.
As an in-depth discussion would go beyond the scope of this paper, we are going to present the results only qualitatively. Details can be found in \cite{Reinhart_2019,Alex_2020_PhD}.

\subsection{Metric cosmology}\label{cosmo}
We first apply the steps outlined in Algorithm \ref{Algo1} to a spatially homogeneous and isotropic metric
spacetime $(M, g)$ (see \cite{Wald_1984}).
More precisely, we consider a metric spacetime where there exists a diffeomorphism $\phi \colon \mathbb{R} \times \Sigma \rightarrow M$, with induced 1-parameter family of embeddings $\phi_{\lambda} \colon \Sigma \rightarrow M$, $\phi_{\lambda}(s)\vcentcolon=\phi(\lambda,s)$,
\begin{equation}
g(\partial_t, \partial_t) = 1 \quad  \text{and} \quad \mathrm dt(X) = 0 \implies  g(\partial_t, X) = 0,   
\end{equation}
with $t \vcentcolon= \pi_{\mathbb{R}} \circ \phi^{-1}$ and moreover,
for all $\lambda \in \mathbb{R}$, the metric 3-manifold $(\Sigma, \gamma_{\lambda}\vcentcolon= \phi_{\lambda}^{\ast} g)$ is homogeneous and isotropic. We define the \emph{scale factor} $a$ as
\begin{equation}
    a(\lambda) \vcentcolon= \sqrt{-\operatorname{det}(\gamma_{\lambda})}^\frac{1}{3}.
\end{equation}
Because all information about the cosmological metric spacetime is encoded in $\partial_t$ and $a$, we construct the gravitational dynamics over the \emph{cosmological bundle} $F_C = TM \oplus_M \mathrm{Vol}^{\frac{1}{3}}(M)$ of vectors and densities of weight $\frac{1}{3}$. The vertical coefficients of vector fields are
\begin{equation}
    C^{a m}_{b n} = \delta^{a}_{n} \delta^{m}_{b}
\end{equation}
and the vertical coefficients of $\frac{1}{3}$ densities are
\begin{equation}
    C^{m}_{n} = -\frac{1}{3} \delta^{m}_{n}.
\end{equation}
Setting up and solving Eq.~(\ref{DiffeoEqn}) yields, up to boundary terms, the action
\begin{equation}\label{CosmoAction}
    S_\text{grav}[a] =
    \frac{1}{2\kappa} \int a^3 \left [-6 \left(\frac{\dot a}{a}\right)^2 -4 U^a_{,a}\frac{\dot a}{a} - \frac{2}{3}\left(U^a_{,a}\right)^2 - 2 \Lambda \right]\mathrm d^4x,
\end{equation}
with $U \vcentcolon= \partial_t$ and $\dot a \vcentcolon= \partial_t a$.
Variations of (\ref{CosmoAction}) w.r.t.~$a$ and $\partial_t$ reproduce the well-known \emph{Friedmann equations}, which read in coordinates where $U^a = const$
\begin{equation}
  \begin{aligned}
    \left(\frac{\dot a}{a}\right)^2 - \frac{\Lambda}{3} = {} & \frac{\kappa}{3} \rho, \\
    \frac{\ddot a}{a} - \frac{\Lambda}{3} = {} & -\frac{\kappa}{6}\left(\rho + 3 p\right).
  \end{aligned}
\end{equation}
Here we introduced the \emph{energy density}
\begin{equation} \label{energy}
    \rho = \frac{1}{a^3}\left\lbrack -\frac{a}{3} \frac{\delta (a^3L_\text{matter})}{\delta a} + U^p \frac{\delta (a^3L_\text{matter})}{\delta U^p}\right\rbrack
\end{equation}
and the \emph{pressure}
\begin{equation}\label{pressure}
    p = \frac{1}{a^3}\left\lbrack \frac{a}{3} \frac{\delta (a^3L_\text{matter})}{\delta a}\right\rbrack.
\end{equation}
These notions of energy density and pressure correspond to the respective notions in general relativity, where both are defined as constituents of the stress-energy tensor of a perfect fluid
\begin{equation}\label{energy-momentum}
    T^{ab} = \frac{2}{\sqrt{-g}} \frac{\delta(\sqrt{- g}L_\text{matter})}{\delta g_{ab}} = (\rho + p)U^a U^b + pg^{a b}.
\end{equation}
This correspondence can be verified for e.g., a spatially homogeneous and isotropic scalar field $\phi$ with action
\begin{equation}
    S_\text{matter}[\phi] = \int\sqrt{-g}\left\lbrack g^{ab} \phi_{,a} \phi_{,b} - V(\phi)\right\rbrack\mathrm d^4x = \int a^3 \left\lbrack (U^a\phi_{,a})^2 - V(\phi) \right\rbrack \mathrm d^4x \nonumber,
\end{equation}
whose energy density and pressure calculated by both methods (either using (\ref{energy}) and (\ref{pressure}) or using (\ref{energy-momentum})) are
\begin{equation}
  \begin{aligned}
    \rho = {} & (\dot \phi)^2 + V(\phi), \\
    p = {} & (\dot \phi)^2 - V(\phi).
  \end{aligned}
\end{equation}

We thus recovered the well-known dynamics of a spatially homogeneous and isotropic metric coupled with such matter, but \emph{without prior knowledge of the Einstein-Hilbert action or the Einstein field equations}, just from the properties of the cosmological bundle $F_C$.

\subsection{Perturbative general relativity}
We apply the perturbative construction manual from Algorithm \ref{Algo2} to the case of gravity being described by a metric tensor field, i.e., a section of $F_{\text{GR}}$. 
Moreover, we couple the metric theory of gravity to a \emph{Klein-Gordon} scalar field:
\begin{equation}\label{KGL}
    \mathcal{L}_{\text{KG}} = \frac{1}{2} \left ( g^{ab} \phi_a \phi_b - m^2 \phi^2\right )\sqrt{-g} \mathrm{d}^4x.
\end{equation}
We choose $N_A = \eta_{ab} J_A^{ab} $ as fiber coordinate value of the Lorentz invariant expansion point. The vertical coefficients of the diffeomorphism Lie algebra action of $F_{\text{GR}}$ are
\begin{equation}
    C^{Am}_{Bn} = -2 I^A_{np} J_B^{mp}.
\end{equation}
The Lorentz invariant expansion coefficients are constructed using computer algebra specifically designed for this task \cite{Reinhart_2019_sparse-tensor}. This Haskell library is capable of computing a Lorentz invariant basis of tensors with given index structure and symmetries as well as setting up and solving tensor equations. The number of arbitrary constants in the
particular expansion coefficients is displayed in Table \ref{GRExp}.
\begin{table}[htp]
\centering 
  \begin{tabularx}{\textwidth}{
   >{\raggedright\arraybackslash}X
   >{\centering\arraybackslash}X
   >{\centering\arraybackslash}X}\toprule\toprule
    Expansion coefficient & Dimension & Constants   \\ \midrule
    $a_0$ & 1 & $\{\mu_1\}$ \\
    $a^A$ & 1 & $\{\mu_2\}$ \\
    $a^{AI}$ & 2 & $\{\nu_1, \nu_2\}$ \\
    $a^{AB}$ & 2 & $\{\mu_3, \mu_4 \} $ \\
    $a^{ApBq}$ & 6 & $\{\nu_3,\dots ,\nu_8\}$ \\
    $a^{ABI}$ & 5 & $\{ \nu_9,\dots ,\nu_{13} \}$ \\
    $a^{ABC}$ & 3 & $\{ \mu_5,\dots \mu_7 \}$\\
    $a^{ABpCq}$ & 21 & $\{\nu_{14},\dots ,\nu_{34} \}$ \\
    $a^{ABCI}$ & 13 & $\{ \nu_{35},\dots ,\nu_{47}\}$\\ \bottomrule\bottomrule
\end{tabularx}
\caption{Dimensions and parameters of the Lorentz invariant expansion coefficients for $\mathcal{L}_{\text{GR}}$.}\label{GRExp}
\end{table}
After inserting the thus obtained expressions in Eqs.~(\ref{order1})--(\ref{order3}) and solving the resulting linear system, we end up with a perturbative Lagrangian that features 2 undetermined parameters, $\mu_1$ and $\nu_1$. 

Moreover, it turns out that the two principal polynomials already coincide in $\mathcal{O}(2)$. 
Thus, already the required diffeomorphism invariance yields the correct causal structure and the construction procedure terminates. 
The obtained result coincides with the perturbative expansion of the Einstein-Hilbert action
\begin{equation}
  S_\text{E.-H.} = \int \frac{1}{2\kappa} (R - 2 \Lambda) \sqrt{-g} \mathrm d^4 x
\end{equation}
around $\eta_{ab}$, with the two remaining constants $\nu_1$ and $\mu_1$ representing the gravitational and cosmological constant.

\subsection{Perturbative area metric gravity}
As a second example we consider a theory of gravity that describes the gravitational field as a $(0,4)$ tensor field that satisfies the symmetries
\begin{equation}\label{areaSym}
    G_{abcd} = -G_{bacd} = G_{cdab}.
\end{equation}
We call this tensor field \emph{area metric} and denote the corresponding vector bundle by $F_{\text{area}}$. 
The area metric is deeply connected to the premetric treatment of electrodynamics, a formulation of classical electrodynamics that does not rely on the geometric background provided by the usual metric tensor field \cite{Hehl_2003,Hehl_2005,L_mmerzahl_2004,Obukhov_1999,Itin_2009}.
In this context one can show that the most \emph{general}, \emph{linear} theory of \emph{electrodynamics} (GLED) that features the conservation of electric charges and satisfies the Lorentz force law is given by the Lagrangian\footnote{Here, the inverse of the area metric is implicitly defined as $G^{a b p q} G_{p q c d} = 4 \delta^{[a}_{\vphantom{d}c} \delta^{b]}_{d}$.}
\begin{equation}
    \mathcal{L}_{\text{GLED}} =  G^{abcd}F_{ab}F_{cd}\,\omega_G\mathrm{d}^4x,
\end{equation}
where $F_{ab}$ is the usual electromagnetic field strength 2-form and $\omega_G$ is any density constructed from the area metric, for instance, $\omega_G = G_{abcd}\epsilon^{abcd}$. 
In particular, for the special case of
\begin{equation}
    G^{abcd} = 2 g^{c[a}g^{b]d} \quad \text{and} \quad \omega_{G}=\sqrt{-g},
\end{equation}
the GLED Lagrangian reproduces standard Maxwell electrodynamics on a metric background $g_{ab}$.

Note that $F_{\text{area}}$ possesses 21-dimensional fibers. Thus, the area metric constitutes a much richer structure than the usual metric tensor field. This is also reflected in the number of \emph{area metric curvature invariants}, which using Theorem \ref{FormalSol} is computed as
\begin{equation}
\begin{aligned}
  k = {} & \operatorname{dim}(J^2F) - 140 \\
    = {} & 4 + 21 + 21\cdot 4 + 21\cdot10 - 140 \\
    = {} & 179.
\end{aligned}
\end{equation}
Therefore, we also expect the perturbative theory of area metric gravity to contain more unknown parameters.

Following the steps outlined in Algorithm \ref{Algo2}
we are going to construct a third-order expansion of area metric gravity.
We choose
\begin{equation}
N_A = J_A^{a b c d} \left(2 \eta_{ac} \eta_{bd} - \epsilon_{abcd}\right),
\end{equation}
because at this expansion point we recover Maxwell electrodynamics on Minkowski spacetime from $\mathcal{L}_{\text{GLED}}$.
Thus, we can interpret the perturbative expansion of area metric gravity as being performed around Minkowski spacetime.
The vertical coefficients are
\begin{equation}\label{areaGotayMInter}
    C_{An}^{Bm} = -4 I^B_{nbcd} J_A^{mbcd}.
\end{equation}
The Lorentz invariant expansion coefficients of the power series Lagrangian (\ref{LperRed}) are again computed with computer algebra \cite{Reinhart_2019_sparse-tensor} and are displayed in Table \ref{AreaExp}.
\begin{table}[htp]
\centering 
  \begin{tabularx}{\textwidth}{
   >{\raggedright\arraybackslash}X
   >{\centering\arraybackslash}X
   >{\centering\arraybackslash}X}\toprule\toprule
    Expansion coefficient & Dimension & Constants   \\ \midrule
    $a_0$ & 1 & $\{\mu_1\}$ \\
    $a^A$ & 2 & $\{\mu_2,\mu_3\}$ \\
    $a^{AI}$ & 3 & $\{\nu_1,\dots , \nu_3\}$ \\
    $a^{AB}$ & 6 & $\{\mu_4,\dots , \mu_9 \} $ \\
    $a^{ApBq}$ & 15 & $\{\nu_4,\dots ,\nu_{18}\}$ \\
    $a^{ABI}$ & 16 & $\{ \nu_{19},\dots ,\nu_{34} \}$ \\
    $a^{ABC}$ & 15 & $\{ \mu_{10},\dots \mu_{24} \}$\\
    $a^{ABpCq}$ & 110 & $\{\nu_{35},\dots ,\nu_{144} \}$ \\
    $a^{ABCI}$ & 72 & $\{ \nu_{145},\dots ,\nu_{216}\}$ \\ \bottomrule\bottomrule
\end{tabularx}
\caption{Dimensions and parameters of the Lorentz invariant expansion coefficients for $\mathcal{L}_{\text{Area}}$.}\label{AreaExp}
\end{table}
Now we obtain a total of 240 such constants. Plugging the expansion coefficients into Eqs.~(\ref{order1})--(\ref{order3}) and solving the corresponding linear system, the number of parameters is reduced to 52. 10 of these parameters occur in expressions that do not involve derivatives such as $H_{Ap}$, $H_{AI}$, i.e., are of type $\mu$, while 42 occur in expressions that do involve derivatives and thus are denoted as $\nu$.

The matter principal polynomial of GLED was first computed by Rubilar \cite{Itin_2009}.
Computing the gravitational principal polynomial of the perturbative expansion of area metric gravity, we finally see that, up to our chosen perturbation order, the two polynomials describe precisely the same vanishing set and thus, in particular, satisfy the second condition (\ref{pertA2}). 
Thus, also for the case of perturbative area metric gravity, already the imposed diffeomorphism invariance renders the two theories causally compatible and the construction algorithm terminates. 
We therefore find that the most general theory of gravity that is compatible with a linear theory of electrodynamics contains 52 unknown parameters in its cubic expansion. The precise expression that we obtained is provided in \cite{Reinhart_2019}.

\section{Conclusions}
We translated the two fundamental requirements \ref{axioms_intro_1} and \ref{axioms_intro_2} for any alternative theory of gravity into rigorous mathematics and examined how these two requirements guide the perturbative as well as the nonperturbative construction of theories. The obtained results have been formulated as a concise step-by-step manual for explicitly computing the most general dynamical laws consistent with the two fundamental requirements that govern any tensorial gravitational field at wish.

We have further tested the two construction manuals by considering metric theories of gravity both perturbatively and under the assumption of a cosmological symmetry. In both cases the corresponding theory of gravity that follows from the Einstein-Hilbert Lagrangian has been recovered, without having ever included information about the dynamics of general relativity. Consequently, the first two tests are considered a success.

Moreover, we have constructed the third-order perturbative Lagrangian of the most comprehensive theory of gravity that is consistent with general linear electrodynamics. Such a third-order Lagrangian for the first time enables the prediction of gravitational wave emission in this highly significant alternative theory of gravity and thus can ultimately be used to put the physically well-motivated area metric to the test. This is a considerable improvement over the previous situation, where only a second-order Lagrangian has been reliably derived by means of canonical closure \cite{Schneider_2017}. With the emission of gravitational waves from a gravitationally bound system being an effect of second order in the field equations, only the emission of waves from nongravitationally bound systems has been studied so far \cite{M_ller_2018}. Using our result, which extends one order higher, to overcome this obstacle is precisely what we consider as one of the main interests of future research that should build up on the foundation laid out by this paper.  

%\bibliography{bibliography}

\end{document}